\pdfoutput=1
\documentclass[11pt]{article}
\usepackage[T1]{fontenc}
\usepackage[utf8]{inputenc}
\usepackage{lmodern}
\usepackage{microtype}
\usepackage[a4paper,margin=1in]{geometry}
\usepackage{amsmath,amssymb}
\usepackage{graphicx}
\usepackage{booktabs}
\usepackage{longtable}
\usepackage{array}
\usepackage{enumitem}
\usepackage[font=small,labelfont=bf]{caption}
\usepackage[authoryear,round]{natbib}
\usepackage{xcolor}
\usepackage{float}
\usepackage{placeins}
\usepackage{url}
\usepackage[colorlinks=true,linkcolor=blue!50!black,citecolor=blue!50!black,urlcolor=blue!50!black,breaklinks=true]{hyperref}
\setlist{itemsep=2pt,topsep=4pt}

\graphicspath{{figures/}}
\hypersetup{pdftitle={Show Your Work: Can Small Open Models Read SEC Filings with Evidence a Machine Can Check?},pdfauthor={Krishna Nandula}}

\title{Show Your Work: Can Small Open Models Read SEC Filings with Evidence a Machine Can Check?\thanks{Preprint. This report has not been peer-reviewed.}}
\author{Krishna Nandula\\Independent Researcher\\\texttt{krsna.nandula@gmail.com}}
\date{November 2026}

\begin{document}
\maketitle

\begin{abstract}
A fluent answer from a language model is not a verified one. We ask whether small open models that run on a single consumer GPU can extract numbers from SEC filings \emph{with evidence a machine can check}: the page and a verbatim quote containing the number. We introduce Filing-Evidence, a benchmark of 1,000 questions over 90 recent 10-K and 10-Q filings of 45 US companies, with answer keys taken from the filers' inline XBRL tags. We evaluate ten instruct models from ten developers (7--21B parameters) on one RTX~4090 in three setups: plain, check-and-retry, and the model's own reasoning mode. Accuracy with checkable evidence ranges from 17.1\% to 82.7\% in the plain setup, and the best result is 84.5\% (Qwen3.5-9B). An automatic checker rejects many wrong answers, but between 16 and 167 answers per 1,000 give a wrong value for an answerable question and still pass every check. In the plain setup, 57\% of these are scale or unit errors and 30\% take another period or column of the right line. Reasoning significantly reduces such errors for 4 of 5 models that offer it. Retrying after a failed check raises accuracy by up to 3.9 points but never reduces them. Questions, model outputs, code and protocol are public.
\end{abstract}


\section{Introduction}
An analyst who copies a number from a filing can show where it came from. For a language model to be trusted with the same task, its answers should be auditable: each number should come with evidence that software can verify. This paper measures how far small open models are from that standard when they must name the page and quote the exact line containing the number.

We study models that fit on one 24\,GB consumer GPU, because an organisation can run them on its own hardware and documents. Three research questions were fixed in a written protocol before the main runs:
\begin{description}
\item[RQ1.] How reliably do these models extract a requested number with automatically checkable evidence, and answer ``not found'' when the number is absent from the pages provided?
\item[RQ2.] In which kinds of filing content, and with which kinds of error, do they fail?
\item[RQ3.] How much do two training-free interventions help, automatic checking with one retry and the model's own reasoning mode, and at what cost?
\end{description}
Our contributions are: (1) a benchmark whose answer keys come from filers' machine-readable tags; (2) a model-independent checker that separates wrong answers it can reject from wrong answers that pass; (3) a paired comparison of ten open models in up to three setups; and (4) an analysis showing that most errors passing the checker are misread units or columns on the correct line.

\section{Related work}
Financial question answering has been studied with FinQA \citep{chen2021finqa}, TAT-QA \citep{zhu2021tatqa}, FinanceBench \citep{islam2023financebench} and DocFinQA \citep{reddy2024docfinqa}. These benchmarks focus mainly on answer correctness and numerical reasoning. Work on attribution evaluates whether generated statements are supported by cited sources \citep{bohnet2022attributed,gao2023alce,liu2023verifiability}. Filing-Evidence combines the two concerns in a narrow, fully automatic setting: every answer must carry a verbatim quote, and correctness is graded against inline XBRL \citep{xbrl2013inline}.

\section{Benchmark design}
\begin{figure}[!htbp]
\centering\includegraphics[width=\linewidth]{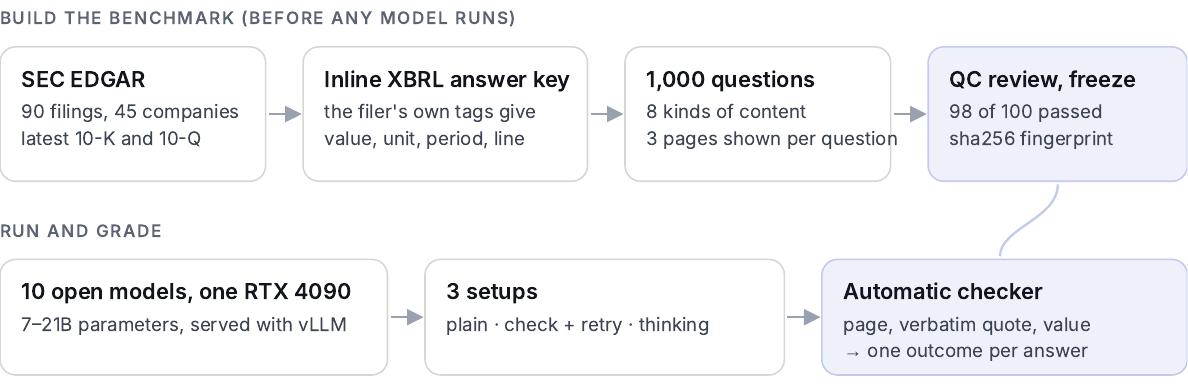}
\caption{Construction and evaluation pipeline.}\label{fig:pipeline}
\end{figure}

\paragraph{Filings and answer key.} We used the most recent 10-K and 10-Q filed on or before 15 November 2026 by 45 large US companies, 90 filings in total (89 filed in 2026), obtained from SEC EDGAR \citep{sec_edgar}. Companies span technology, banking, insurance, REITs, utilities, healthcare, consumer, industrial and other sectors. Gold answers come from inline XBRL facts displayed on a page, with their value, unit, scale, period, dimensions and evidence line.

\paragraph{Questions.} Each question names an item, company and period and is shown with three pages: the answer page, an adjacent page and a random distractor, in random order. The benchmark has eight slices of 125 questions (Table~\ref{tab:slices}). In S8 the requested figure is absent from the pages and the correct response is ``not found''.

\begin{table}[!htbp]\centering\small
\caption{Benchmark slices (125 questions each).}\label{tab:slices}
\begin{tabular}{llr}\toprule
Slice & Content & Companies\\\midrule
S1 & Main statements, current period & 44\\
S2 & Main statements, prior period & 44\\
S3 & 10-Q quarter vs year-to-date & 36\\
S4 & Notes tables & 44\\
S5 & Segments & 29\\
S6 & Numbers in prose & 43\\
S7 & Units and per-share & 42\\
S8 & Not in the provided pages & 45\\
\bottomrule\end{tabular}
\end{table}

\paragraph{Quality control.} Before the main runs, stratified samples of 100 questions were reviewed against the rendered filing. The review was performed by an AI assistant under the author's direction. Rounds one to three scored 87, 76 and 96 of 100, failing on question wording; the builder was revised after each. Round four scored 98 of 100, meeting the protocol's threshold of 98, and the dataset was frozen with a SHA-256 fingerprint (\texttt{4fa8f252292167fb\ldots}). All 400 reviewed gold values were judged correct. The author did not independently verify the full sample (Section~\ref{sec:limits}).

\begin{figure}[!htbp]
\centering\includegraphics[width=\linewidth]{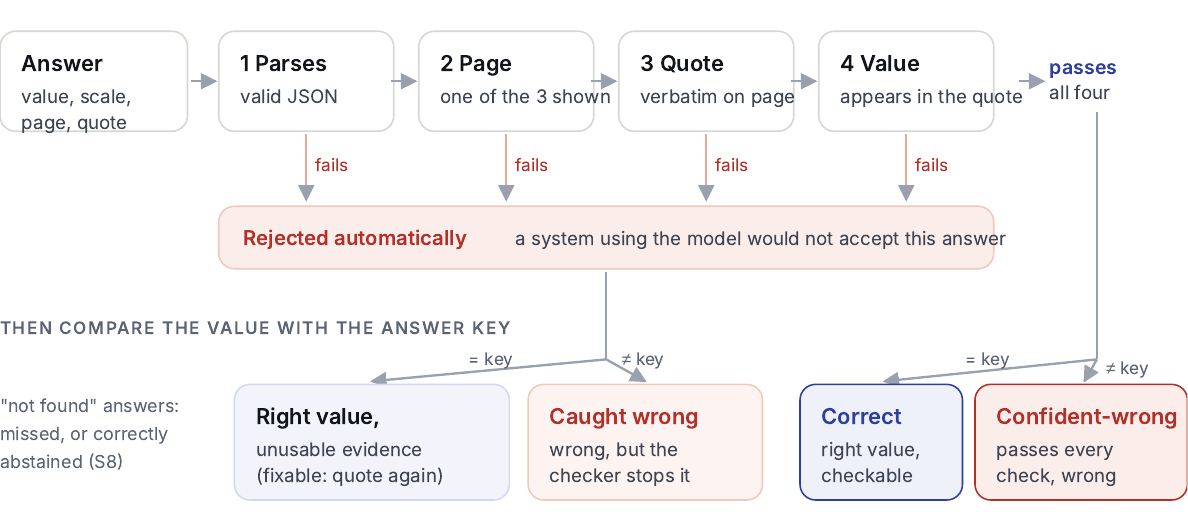}
\caption{The checker. It uses the answer and the pages only; comparison with the answer key happens afterwards, for scoring.}\label{fig:checker}
\end{figure}

\paragraph{Checker and metrics.} Models return JSON with a value, scale, page and quoted line, or \texttt{not\_found}. An answer passes when it parses, cites a page that was shown, quotes text verbatim from that page, and the value appears in the quote (Figure~\ref{fig:checker}). The prompt asks for the complete line, but the checker accepts any contiguous passage after normalising whitespace: 923 of the 5{,}293 correct plain-setup answers across the ten models quote something other than one whole line. Sign is not graded because filers display negative values inconsistently. \emph{Accuracy} is the share of the 875 answerable questions answered correctly with passing evidence. The \emph{confident-wrong rate} is the share of all 1,000 answers that give a wrong value for an answerable question and pass every check. Answers that invent a figure on S8 are counted separately, as failed abstentions. \emph{Abstention accuracy} is the share of S8 questions answered ``not found''. Rates carry 95\% Wilson intervals \citep{wilson1927}. Paired comparisons use exact McNemar tests \citep{mcnemar1947} with Holm correction \citep{holm1979} at $\alpha=0.05$. For each metric, the 23 within-model setup comparisons form one family (Appendix~\ref{app:within}), and the pairwise model comparisons within each setup form another.

\FloatBarrier
\section{Models and setups}
\begin{figure}[!htbp]
\centering\includegraphics[width=\linewidth]{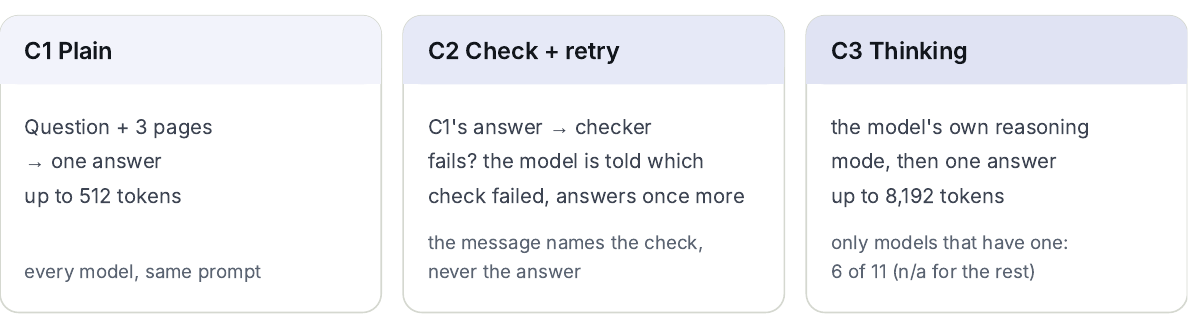}
\caption{The three experimental setups.}\label{fig:setups}
\end{figure}

All models were served with vLLM~0.33.0 \citep{kwon2023vllm} on one NVIDIA RTX~4090 at temperature~0, using each model's chat template and one shared prompt (Table~\ref{tab:models}). Qwen3-8B, the model of an earlier pilot, is a supplementary baseline and is excluded from model rankings. The setups are shown in Figure~\ref{fig:setups}. C2 reuses C1's first answer and retries only answers that fail a check; the retry message names the failed check and never reveals the answer. gpt-oss-20b has no non-reasoning mode, so its C1 and C2 use low reasoning effort and its C3 uses high effort.

\begin{table}[!htbp]\centering\footnotesize
\caption{Models evaluated. Parameter counts are as published and may include unused vision components. * Supplementary baseline.}\label{tab:models}
\setlength{\tabcolsep}{4pt}\begin{tabular}{lllllll}\toprule
Model & Developer & Params & Precision & Created & Reasoning & Licence\\\midrule
Qwen3.5-9B & Alibaba & 9.7B & bf16 & 2026-02 & yes & Apache 2.0\\
Granite-4.2-8B & IBM & 8.8B & bf16 & 2026-08 & yes & Apache 2.0\\
Ministral-3-8B & Mistral & 8.9B & bf16 & 2025-10 & no & Apache 2.0\\
OLMo-3-7B & AllenAI & 7.3B & bf16 & 2025-11 & no & Apache 2.0\\
Nemotron-Nano-9B-v2 & NVIDIA & 8.9B & bf16 & 2025-08 & yes & NVIDIA Open Model\\
gpt-oss-20b & OpenAI & 20.9B & mxfp4 & 2025-08 & yes & Apache 2.0\\
Phi-4 & Microsoft & 14.7B & fp8 & 2024-12 & no & MIT\\
Gemma-3-12B & Google & 12.2B & fp8 & 2025-03 & no & Gemma Terms\\
GLM-4.6V-Flash & Zhipu (Z.ai) & 10.3B & bf16 & 2025-12 & yes & MIT\\
Llama-3.1-8B & Meta & 8.0B & bf16 & 2024-07 & no & Llama 3.1\\
Qwen3-8B* & Alibaba & 8.2B & bf16 & 2025-04 & yes & Apache 2.0\\
\bottomrule\end{tabular}
\end{table}

\FloatBarrier
\section{Results}
\subsection{RQ1: Reliability}
\begin{figure}[!htbp]
\centering\includegraphics[width=.86\linewidth]{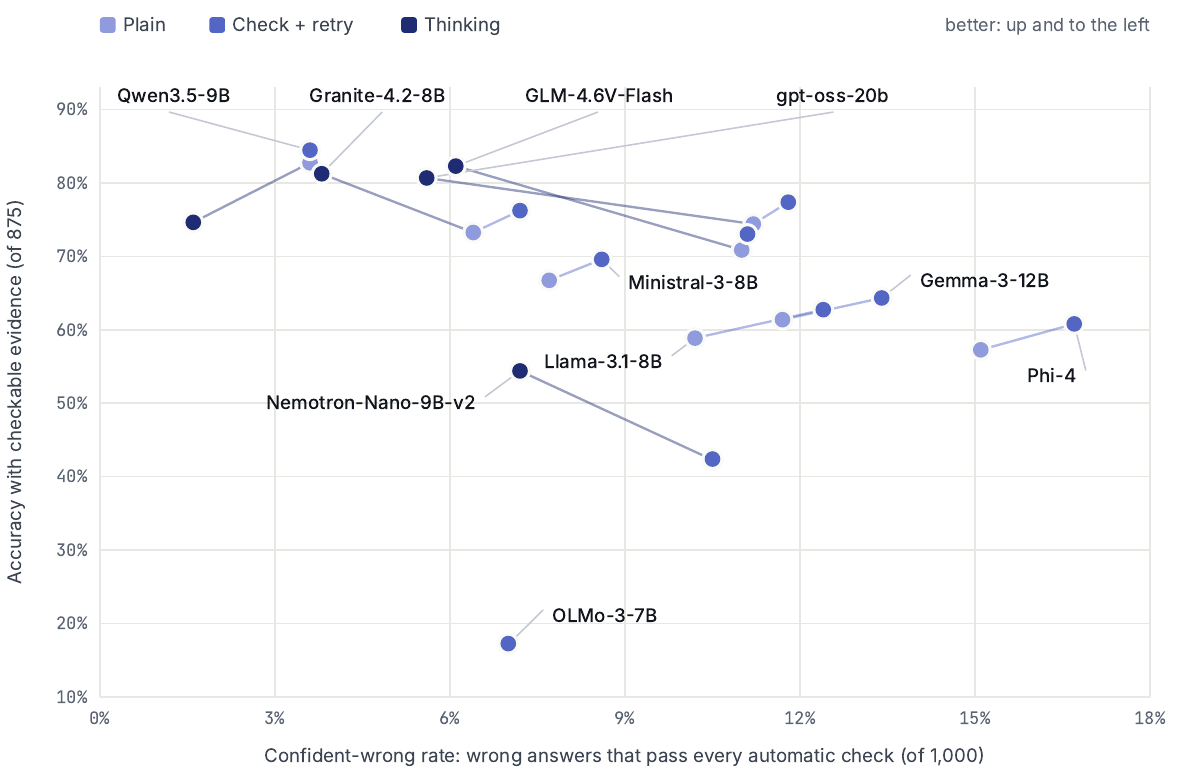}
\caption{Accuracy against confident-wrong rate for each model and setup. Up and to the left is better.}\label{fig:frontier}
\end{figure}

\begin{figure}[!htbp]
\centering\includegraphics[width=.6\linewidth]{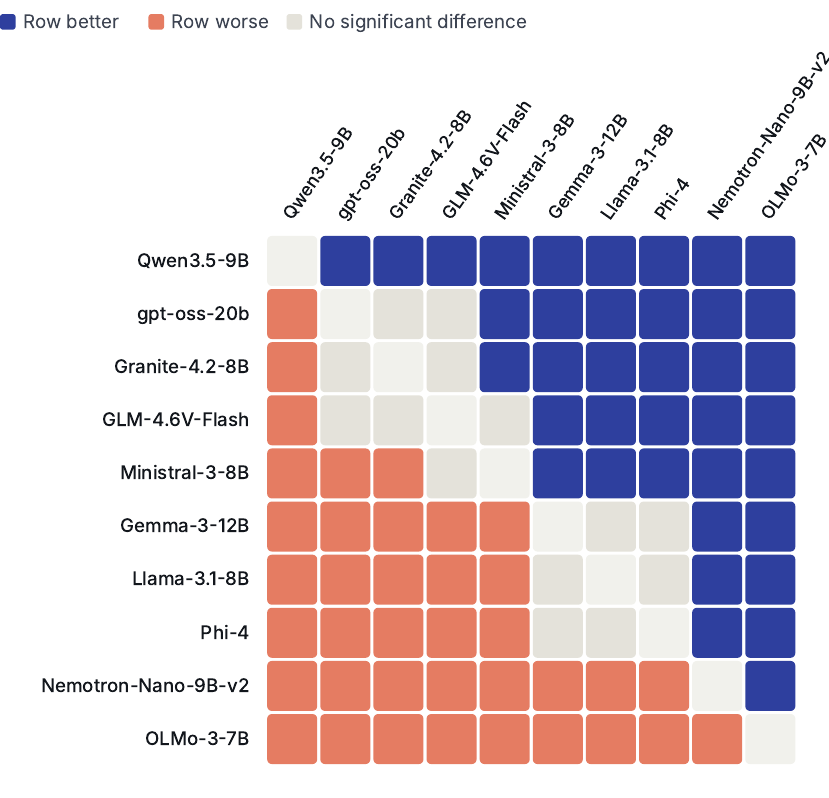}
\caption{Pairwise differences in plain-setup accuracy (exact McNemar, Holm-corrected over 45 pairs).}\label{fig:significance}
\end{figure}

\begin{figure}[!htbp]
\centering\includegraphics[width=.86\linewidth]{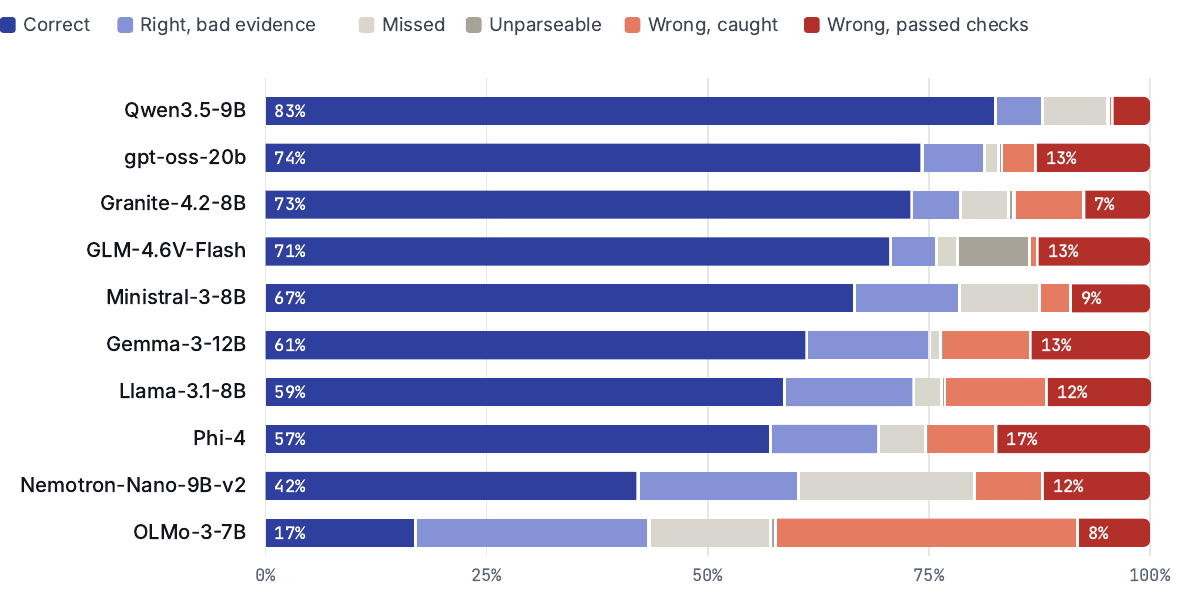}
\caption{Outcomes of the 875 answerable questions in the plain setup.}\label{fig:outcomes}
\end{figure}

Table~\ref{tab:results} and Figure~\ref{fig:frontier} summarise the results. In the plain setup Qwen3.5-9B reaches 82.7\% (95\% CI 80.1--85.1\%) and is significantly more accurate than every other model (Figure~\ref{fig:significance}). Figure~\ref{fig:outcomes} shows where the answers go. Abstention accuracy ranges from 77.6\% to 100.0\%. Across the ten models, 73 invented answers to S8 questions passed every check in the plain setup (0 to 24 per model); these are not included in the confident-wrong counts.

\begin{table}[!htbp]\centering\footnotesize
\caption{Accuracy (of 875), confident-wrong answers (of 1,000) and abstention (of 125, plain setup). ``--'': no reasoning mode. * Supplementary baseline.}\label{tab:results}
\begin{tabular}{lrrrrrrr}\toprule
& \multicolumn{2}{c}{C1 Plain} & \multicolumn{2}{c}{C2 Retry} & \multicolumn{2}{c}{C3 Thinking} & \\
Model & Acc. & CW & Acc. & CW & Acc. & CW & Abst.\\\midrule
Qwen3.5-9B & 82.7\% & 36 & 84.5\% & 36 & 74.6\% & 16 & 124\\
gpt-oss-20b & 74.4\% & 112 & 77.4\% & 118 & 80.7\% & 56 & 112\\
Granite-4.2-8B & 73.3\% & 64 & 76.2\% & 72 & 81.3\% & 38 & 121\\
GLM-4.6V-Flash & 70.9\% & 110 & 73.0\% & 111 & 82.3\% & 61 & 118\\
Ministral-3-8B & 66.7\% & 77 & 69.6\% & 86 & -- & -- & 124\\
Gemma-3-12B & 61.4\% & 117 & 64.3\% & 134 & -- & -- & 97\\
Llama-3.1-8B & 58.9\% & 102 & 62.7\% & 124 & -- & -- & 110\\
Phi-4 & 57.3\% & 151 & 60.8\% & 167 & -- & -- & 116\\
Nemotron-Nano-9B-v2 & 42.3\% & 105 & 42.4\% & 105 & 54.4\% & 72 & 125\\
OLMo-3-7B & 17.1\% & 70 & 17.3\% & 70 & -- & -- & 121\\
Qwen3-8B* & 66.1\% & 94 & 67.3\% & 101 & 74.1\% & 37 & 88\\
\bottomrule\end{tabular}
\end{table}

\FloatBarrier
\subsection{RQ2: Where models fail}
\begin{figure}[!htbp]
\centering\includegraphics[width=.86\linewidth]{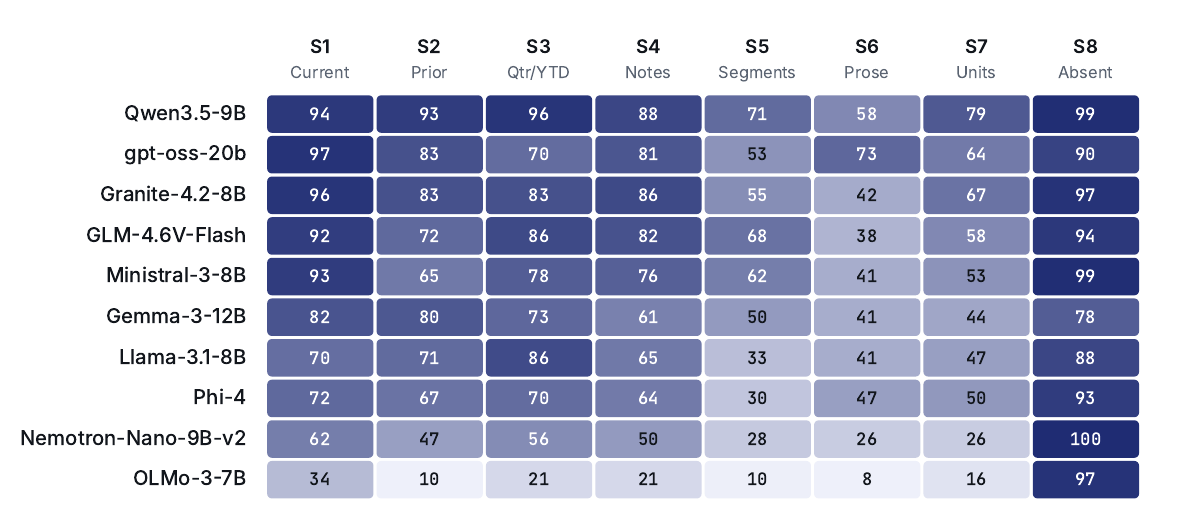}
\caption{Plain-setup accuracy by model and slice (S8: correct ``not found'').}\label{fig:slices}
\end{figure}

Pooled over the ten models, plain-setup accuracy is 79.2\% on current-period statements, but 46.0\% on segments, 41.4\% on numbers in prose and 50.5\% on units and per-share items (Figure~\ref{fig:slices}).

\begin{figure}[!htbp]
\centering\includegraphics[width=.82\linewidth]{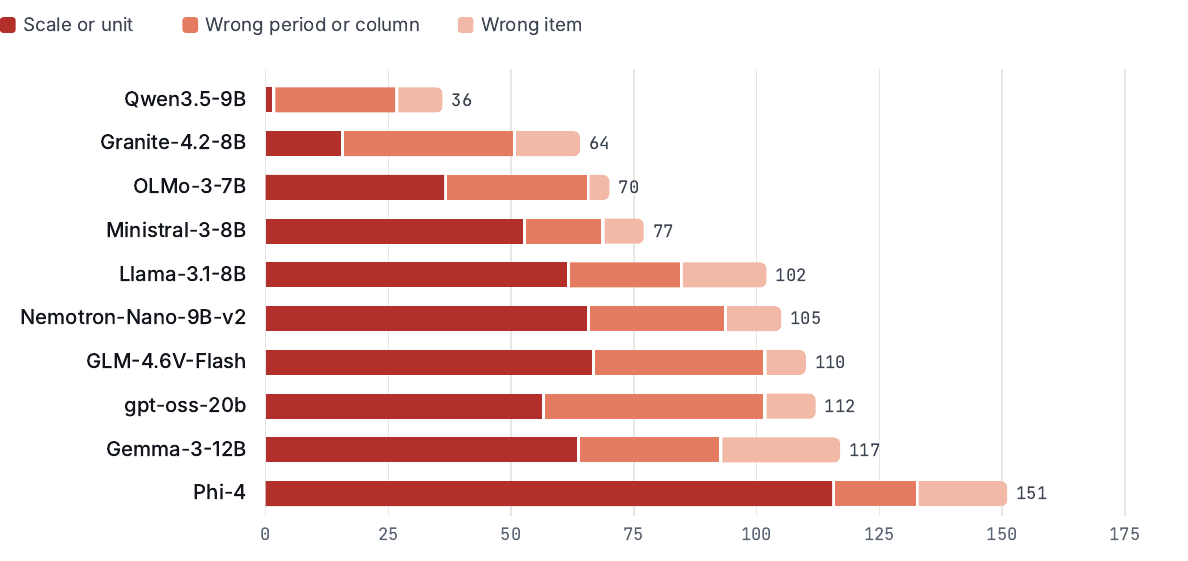}
\caption{Confident-wrong answers by error type, plain setup.}\label{fig:cwmix}
\end{figure}

Error types are assigned by fixed rules in the grading code: the right digits at the wrong scale, another figure from the answer key's evidence line, or otherwise a wrong item. The 944 plain-setup confident-wrong answers across the ten models divide into scale or unit errors (57\%), another period or column of the right line (30\%) and wrong items (13\%), as shown in Figure~\ref{fig:cwmix}. The checker cannot detect the first two classes because the quote is genuine and contains the reported digits. Figure~\ref{fig:example} shows a real case.

\begin{figure}[!htbp]
\small\raggedright
\fbox{\parbox{.97\linewidth}{\raggedright
\textbf{Question.} What was Bank of America Corporation's net change in derivatives for the nine months ended September 30, 2025?\\[3pt]
\textbf{Quoted line.} \texttt{Net change in derivatives (1,496) 636 (2,874) 3,145}\\[3pt]
\textbf{Model answer.} -2874 (in millions), page 48.\quad \textbf{Answer key.} 3,145\\[3pt]
\textbf{Checks.} Parses, page shown, quote verbatim and value in quote all pass.\quad \textbf{Outcome.} Confident-wrong.}}
\caption{A confident-wrong answer from Qwen3.5-9B in the plain setup (Bank of America Corporation, 10-Q). The line is correct but the figure comes from another column.}\label{fig:example}
\end{figure}

\FloatBarrier
\subsection{RQ3: Retry and reasoning}
\begin{figure}[!htbp]
\centering\includegraphics[width=.86\linewidth]{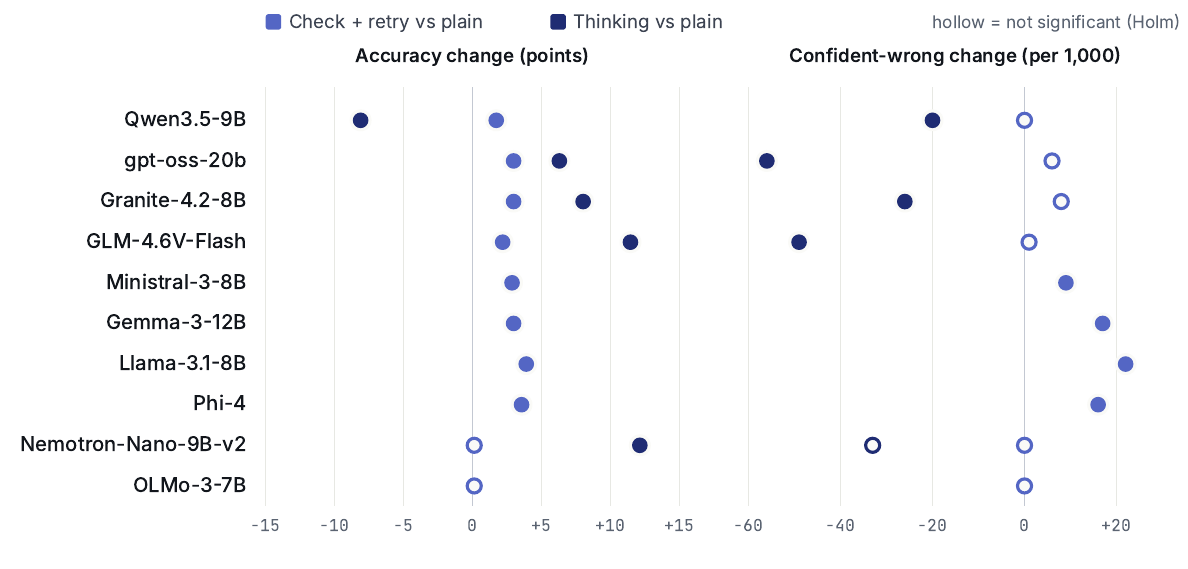}
\caption{Change from the plain setup. Filled markers are significant after Holm correction.}\label{fig:effects}
\end{figure}

Figure~\ref{fig:effects} shows the change from the plain setup. Check-and-retry raises accuracy by 0.1 to 3.9 points, significantly for 8 of the ten models. It never reduces confident-wrong answers, and it significantly increases them for 4 models (Ministral-3-8B, Phi-4, Gemma-3-12B, Llama-3.1-8B). A retry can make a wrong value pass by finding a line that contains it. For Phi-4, 16 of 177 retried answers became confident-wrong.

\begin{figure}[!htbp]
\centering\includegraphics[width=.8\linewidth]{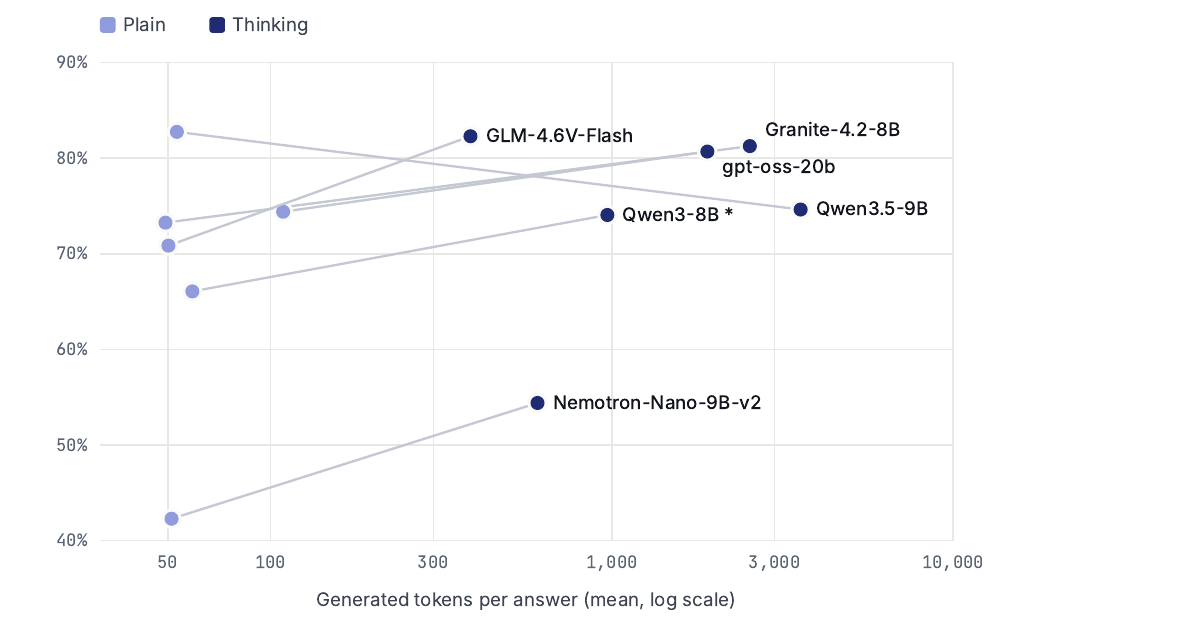}
\caption{Accuracy against mean generated tokens per answer (log scale). * Supplementary baseline.}\label{fig:cost}
\end{figure}

Reasoning reduces confident-wrong answers for all 5 ranked models with a reasoning mode (4 significantly), and it raises accuracy for 4 of them. Qwen3.5-9B is the exception: accuracy falls from 82.7\% to 74.6\%. It reached the pre-specified 8,192-token limit on 180 questions, 88 of which produced no parseable answer; on the 699 answerable questions it finished within the limit, its accuracy was 85.1\%, against 84.4\% for the plain setup on the same questions. Reasoning multiplies mean generated tokens by 8 to 67 times (Figure~\ref{fig:cost}).

\FloatBarrier
\section{Discussion}
Requiring a verbatim quote exposes many errors at low cost, but it does not establish that the selected figure or its units are right. Because most passing errors occur on the correct line, a stronger checker could require the column header and units statement to be quoted and verify their relationship to the selected figure. Retrying should also be treated cautiously: a retried answer that passes may have been adapted to the checker rather than corrected.

\section{Limitations}\label{sec:limits}
\begin{itemize}
\item \textbf{AI-assisted quality control.} The dataset QC review was performed by an AI assistant under the author's direction, and the author did not independently verify the full sample. The planned manual review of 20 error categorisations per model was not carried out.
\item \textbf{Protocol timing.} The protocol was maintained in version control and was not lodged with an external registry. Repository history was later rewritten to edit commit metadata, so its timestamps are not independent evidence of timing. Deviations are logged in the protocol. The protocol's coverage-at-zero-confident-wrong metric was not computed, because every model and setup produced confident-wrong answers and the models report no confidence score to threshold.
\item \textbf{Configuration on benchmark questions.} Serving settings were checked on 24 questions from the benchmark before the main runs.
\item \textbf{Units stated off-page.} After the results, we identified 30 answerable questions whose evidence page shows no units statement, meaning no plural scale word such as ``millions'' and no scale written with the figure. Excluding them changes no model's accuracy by more than 2.3 points (Appendix~\ref{app:sens}) and lowers the scale or unit share of plain-setup confident-wrong answers from 57\% to 54\%.
\item \textbf{Scope.} Results reflect one prompt, one run per configuration, three supplied pages per question and text from large US filers. Model versions and quantisation may affect results.
\item \textbf{Reasoning budget.} The 8,192-token limit truncated some reasoning runs.
\item \textbf{Contamination.} Models may have seen earlier filings during training, although a memorised answer must still be supported by a verbatim quote.
\end{itemize}

\section{Conclusion}
Small open models can often extract figures from SEC filings with verifiable quotes, and model choice matters substantially. Verbatim evidence is necessary but insufficient: errors in units and columns can pass automatic checks. Future work should verify the relationship between a figure, its column and its units.

\FloatBarrier
\section*{Data and code availability}
\noindent Code and protocol:\\\url{https://github.com/pramana-research/filing-evidence}\\[4pt]
Questions, page texts, model outputs and server logs:\\\url{https://huggingface.co/datasets/pramana-research/filing-evidence}

\section*{Acknowledgements and AI use}
The author conceived and directed this study, defined its research questions and design, and made all decisions on scope, methods and publication. Claude (Anthropic) was used as an assistant for implementing code, running analyses, reviewing the quality-control samples and drafting text. The author reviewed the manuscript and takes full responsibility for its content. This work was done independently; it does not represent the views of any employer, and it is not financial or investment advice.

\clearpage
\appendix
\section{Sensitivity to off-page units}\label{app:sens}
\begin{table}[H]\centering\footnotesize
\caption{Accuracy on all 875 answerable questions and after excluding 30 questions without an on-page units statement.}
\begin{tabular}{lrrrrrr}\toprule
Model & C1 all & C1 excl. & C2 all & C2 excl. & C3 all & C3 excl.\\\midrule
Qwen3.5-9B & 82.7\% & 82.7\% & 84.5\% & 84.5\% & 74.6\% & 76.1\%\\
gpt-oss-20b & 74.4\% & 76.0\% & 77.4\% & 79.1\% & 80.7\% & 82.4\%\\
Granite-4.2-8B & 73.3\% & 74.4\% & 76.2\% & 77.3\% & 81.3\% & 83.6\%\\
GLM-4.6V-Flash & 70.9\% & 72.3\% & 73.0\% & 74.6\% & 82.3\% & 83.0\%\\
Ministral-3-8B & 66.7\% & 68.8\% & 69.6\% & 71.7\% & -- & --\\
Gemma-3-12B & 61.4\% & 63.4\% & 64.3\% & 66.5\% & -- & --\\
Llama-3.1-8B & 58.9\% & 60.4\% & 62.7\% & 64.4\% & -- & --\\
Phi-4 & 57.3\% & 58.9\% & 60.8\% & 62.5\% & -- & --\\
Nemotron-Nano-9B-v2 & 42.3\% & 43.8\% & 42.4\% & 43.9\% & 54.4\% & 56.0\%\\
OLMo-3-7B & 17.1\% & 17.4\% & 17.3\% & 17.5\% & -- & --\\
\bottomrule\end{tabular}
\end{table}

\section{Within-model comparisons}\label{app:within}
{\footnotesize
\begin{longtable}{llrrrr}
\caption{Paired exact McNemar tests. ``A/B'' counts questions favouring only the first or second setup; $p$ values are Holm-adjusted.}\\\toprule
Model & Comparison & Acc. A/B & Acc. $p$ & CW A/B & CW $p$\\\midrule\endfirsthead
\toprule Model & Comparison & Acc. A/B & Acc. $p$ & CW A/B & CW $p$\\\midrule\endhead
Qwen3.5-9B & C1 vs C2 & 0/15 & 0.000427 & 0/0 & 1\\
Qwen3.5-9B & C1 vs C3 & 159/88 & 7.35e-05 & 27/7 & 0.0107\\
Qwen3.5-9B & C2 vs C3 & 162/76 & 4.35e-07 & 27/7 & 0.0107\\
Granite-4.2-8B & C1 vs C2 & 0/26 & 4.77e-07 & 0/8 & 0.0579\\
Granite-4.2-8B & C1 vs C3 & 91/161 & 0.00011 & 51/25 & 0.0422\\
Granite-4.2-8B & C2 vs C3 & 96/140 & 0.0201 & 56/22 & 0.00209\\
Ministral-3-8B & C1 vs C2 & 0/25 & 7.75e-07 & 0/9 & 0.0422\\
OLMo-3-7B & C1 vs C2 & 0/1 & 1 & 0/0 & 1\\
Nemotron-Nano-9B-v2 & C1 vs C2 & 0/1 & 1 & 0/0 & 1\\
Nemotron-Nano-9B-v2 & C1 vs C3 & 100/206 & 2.68e-08 & 86/53 & 0.0579\\
Nemotron-Nano-9B-v2 & C2 vs C3 & 101/206 & 3.91e-08 & 86/53 & 0.0579\\
gpt-oss-20b & C1 vs C2 & 0/26 & 4.77e-07 & 0/6 & 0.156\\
gpt-oss-20b & C1 vs C3 & 67/122 & 0.000462 & 76/20 & 1.32e-07\\
gpt-oss-20b & C2 vs C3 & 74/103 & 0.105 & 79/17 & 2.09e-09\\
Phi-4 & C1 vs C2 & 0/31 & 1.96e-08 & 0/16 & 0.000458\\
Gemma-3-12B & C1 vs C2 & 0/26 & 4.77e-07 & 0/17 & 0.000244\\
GLM-4.6V-Flash & C1 vs C2 & 0/19 & 4.2e-05 & 0/1 & 1\\
GLM-4.6V-Flash & C1 vs C3 & 49/149 & 1.45e-11 & 59/10 & 2.62e-08\\
GLM-4.6V-Flash & C2 vs C3 & 52/133 & 4.08e-08 & 60/10 & 1.6e-08\\
Llama-3.1-8B & C1 vs C2 & 0/34 & 2.56e-09 & 0/22 & 8.11e-06\\
Qwen3-8B & C1 vs C2 & 0/11 & 0.00488 & 0/7 & 0.0938\\
Qwen3-8B & C1 vs C3 & 66/136 & 1.14e-05 & 67/10 & 3.73e-10\\
Qwen3-8B & C2 vs C3 & 70/129 & 0.000278 & 73/9 & 3.17e-12\\
\bottomrule
\end{longtable}}

\section{Confident-wrong rates}
\begin{figure}[H]
\centering\includegraphics[width=.86\linewidth]{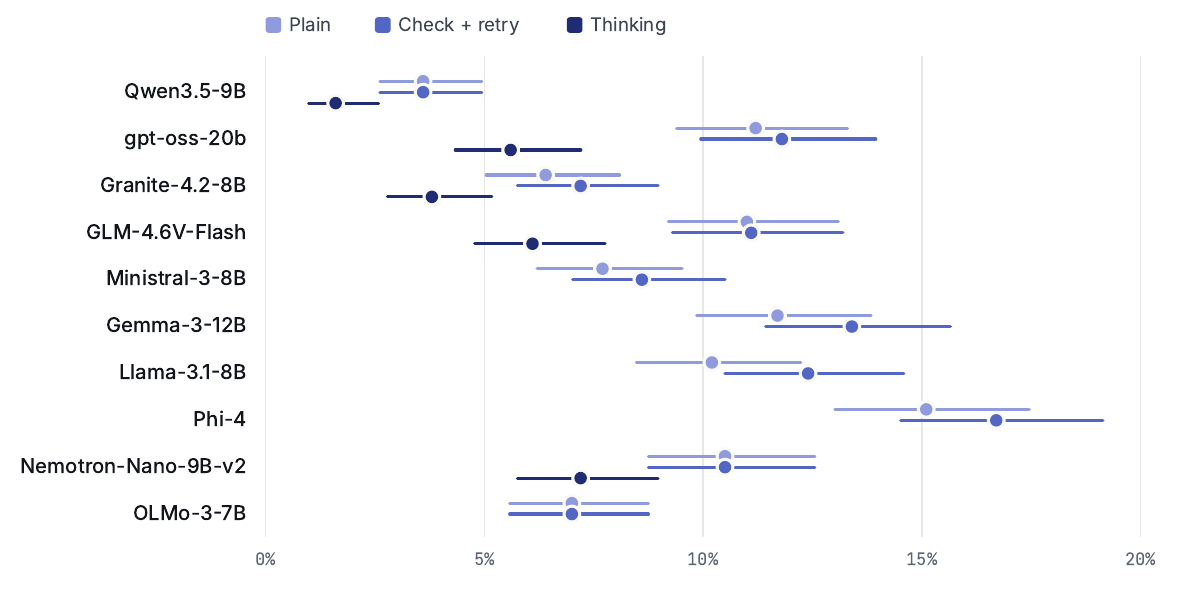}
\caption{Confident-wrong rate with 95\% Wilson intervals.}
\end{figure}


\begin{thebibliography}{13}
\bibitem[Bohnet et~al.(2022)]{bohnet2022attributed} B. Bohnet et al. Attributed question answering: Evaluation and modeling for attributed large language models. \emph{arXiv:2212.08037}, 2022.
\bibitem[Chen et~al.(2021)]{chen2021finqa} Z. Chen et al. FinQA: A dataset of numerical reasoning over financial data. In \emph{Proceedings of EMNLP}, 2021.
\bibitem[Gao et~al.(2023)]{gao2023alce} T. Gao, H. Yen, J. Yu, and D. Chen. Enabling large language models to generate text with citations. In \emph{Proceedings of EMNLP}, 2023.
\bibitem[Holm(1979)]{holm1979} S. Holm. A simple sequentially rejective multiple test procedure. \emph{Scandinavian Journal of Statistics}, 6(2):65--70, 1979.
\bibitem[Islam et~al.(2023)]{islam2023financebench} P. Islam et al. FinanceBench: A new benchmark for financial question answering. \emph{arXiv:2311.11944}, 2023.
\bibitem[Kwon et~al.(2023)]{kwon2023vllm} W. Kwon et al. Efficient memory management for large language model serving with PagedAttention. In \emph{Proceedings of SOSP}, 2023.
\bibitem[Liu et~al.(2023)]{liu2023verifiability} N. F. Liu, T. Zhang, and P. Liang. Evaluating verifiability in generative search engines. In \emph{Findings of EMNLP}, 2023.
\bibitem[McNemar(1947)]{mcnemar1947} Q. McNemar. Note on the sampling error of the difference between correlated proportions or percentages. \emph{Psychometrika}, 12(2):153--157, 1947.
\bibitem[Reddy et~al.(2024)]{reddy2024docfinqa} V. Reddy et al. DocFinQA: A long-context financial reasoning dataset. In \emph{Proceedings of ACL}, 2024.
\bibitem[{SEC}(2026)]{sec_edgar} U.S. Securities and Exchange Commission. EDGAR. \url{https://www.sec.gov/edgar}, 2026.
\bibitem[Wilson(1927)]{wilson1927} E. B. Wilson. Probable inference, the law of succession, and statistical inference. \emph{Journal of the American Statistical Association}, 22(158):209--212, 1927.
\bibitem[{XBRL International}(2013)]{xbrl2013inline} XBRL International. Inline XBRL Part 1: Specification 1.1, 2013.
\bibitem[Zhu et~al.(2021)]{zhu2021tatqa} F. Zhu et al. TAT-QA: A question answering benchmark on a hybrid of tabular and textual content in finance. In \emph{Proceedings of ACL-IJCNLP}, 2021.
\end{thebibliography}
\end{document}